\documentclass[usegraphicx,plainnat]{mn2e}

\title[Fast Optical and X-ray Variability in the UCXB 4U0614+09]{Fast Optical and X-ray Variability in the UCXB 4U0614+09}
\author[P.J. Hakala, P.A. Charles and P. Muhli]{P.J. Hakala$^{1}$, P.A. Charles$^{2}$, P.Muhli$^{3}$ \thanks{E-mail: pahakala@utu.fi} \thanks{Based on observations made with the Nordic Optical Telescope, operated
on the island of La Palma jointly by Denmark, Finland, Iceland,
Norway, and Sweden, in the Spanish Observatorio del Roque de los
Muchachos of the Instituto de Astrofisica de Canarias.}\\
$^{1}$Finnish Centre for Astronomy with ESO (FINCA), V\"ais\"al\"antie 20, University of Turku, FIN-21500, Piikki\"o,
Finland.\\
$^{2}$South African Astronomical Observatory (SAAO), South Africa\\
$^{3}$National Land Survey of Finland, Finland
 }
\begin{document}

\date{Accepted xxxx. Received xxxx; in original form xxxx}

\pagerange{\pageref{firstpage}--\pageref{lastpage}} \pubyear{2011}

\maketitle

\label{firstpage}

\begin{abstract}
We present results from several years of fast optical photometry of 4U0614+091 (V1055 Orionis),
a candidate ultracompact X-ray binary most likely consisting of a neutron star and a 
degenerate secondary. We find evidence for strong accretion-driven variability at all epochs, that manifests itself as red noise. This flickering produces transient peaks in the observed power spectrum in the 15-65 min period range. Only in 
one of our 12 optical datasets can we see evidence for a period that cannot be reproduced using the
red noise model. This period of 51 minutes coincides with the strongest period detected by Shahbaz et al. (2008)
and can thus be taken as the prime candidate for the orbital period of the system. Furthermore, 
we find some tentative evidence for the X-ray vs. optical flux anticorrelation discovered by Machin et al. (1990) using our data together with the all-sky X-ray monitoring data from RXTE/ASM. We propose that the complex 
time series behaviour of 4U0614+09 is a result of drastic 
changes in the accretion disc geometry/structure on time scales from hours to days. Finally we want to draw attention
to the interpretation of moderately strong peaks in the power spectra of, especially accreting, sources. Many
of such ``periods" can probably be attributed to the presence of red noise (i.e. correlated events) in the data.      
 
\end{abstract}


\section{Introduction}

X-ray binaries are systems where either a neutron star or a stellar mass black hole
accretes matter from a ``normal" companion star. If the secondary star is of late spectral
type, the accretion usually takes the form of Roche lobe overflow and an accretion disc 
is formed. These systems are known as low mass X-ray binaries (LMXB's, see for instance Charles \& Coe, 2006
for a review on observational properties of LMXBs) . 

4U0614+09 is an X-ray source first detected by the {\it Uhuru} satellite (Willmore et al., 1974, Forman et al. 1978) and optically identified by Murdin et al. (1974). The source, whilst being a persistent X-ray source, shows variability at a number of timescales (Machin et al. 1990, Shahbaz et al. 2008). Subsequent optical studies (Machin et al. 1990) have revealed a 9.8d period in the optical. This is, however, most likely not the orbital period, but rather a superorbital period related to the precession of, and/or periodic changes in, the accretion disc structure (Machin et al. 1990).  X0614+091 has also shown a type I X-ray burst (Swank et al. 1978) , indicating that the accreting compact object is a neutron star. A recent analysis of burst properties, and in particular the detection of photospheric radius-expansion phase (Kuulkers et al. 2010) has enabled the measurement of distance, 3.2 kpc. Furthermore, 
Strohmayer et al. (2008) detected a 415 Hz pulsation during a type I X-ray burst, which they interpret as the neutron star spin
frequency.   

There are several pieces of evidence supporting the ultracompact nature of the system. Firstly, given the distance estimate of 3.2kpc (Kuulkers et al. 2010), together with an orbital period of 9.8d would lead to a very massive secondary star easily visible in the optical spectra (assuming Roche lobe overflow). There is, however, no sign of the secondary in the spectra. Secondly, the line emission spectrum from the accretion disc lacks all the hallmark emission lines, like the Hydrogen Balmer series and HeII at 4686\AA. (Machin et al., 1990). This, together with the detection of X-ray lines of O, Fe and Ne in the Chandra spectra (Paerels et al., 2001) and detection of C and O lines in the optical (Nelemans et al. 2004) has led to the suggestion that 4U0614+091 is an ultracompact X-ray binary (UCXRB) with a hydrogen and helium deficient donor star i.e. a C/O white dwarf. The direct implication of a WD mass donor is  that the orbital period of the system has to be less than $\sim$ 1 hour. Nelemans et al. (2006) obtained phase resolved spectra over the tentative (unpublished) 50 min period and found a weak sinusoidal radial velocity signal at 48.5min in the line at 4960\AA. However,
the signal was not detected in their period analysis. If real, the amplitude of this variation would imply an inclination of $\sim$ 15 degrees assuming a neutron star mass of 1.4M$\odot$. Schulz et al. (2010) analyzed 200 ksec of Chandra transmission
grating data and find that relativistic effects needed to be included to fit the X-ray spectra. Their fits, the relativistic line profiles in particular, imply that the system inclination should be very high i.e. 86-89 degrees. This is rather contradictory though, since the system does not show any evidence for even partial eclipses, neither in optical nor X-rays.

Another peculiarity associated with the system is that its X-ray emission seems to be
anticorrelated with the optical emission (Machin et al. 1990). Moreover, the X-rays become harder with increasing X-ray luminosity ruling out any photoelectric absorption effects as a cause for changes in the X-ray flux. 

Recently Shahbaz et al. (2008) have published the results of their photometric monitoring programme. They found evidence for three different periods i.e. 42, 51.3 and 64 minutes in their data, with the 51.3 min period showing the clearest case of periodicity in terms of signal-to-noise. Still, the 51.3 min period is only present in some of their data, whilst other observations show evidence for periods at 42.0 and 64.1 minutes. Clearly much more data is required.

\begin{figure*}
\includegraphics[scale=1.0]{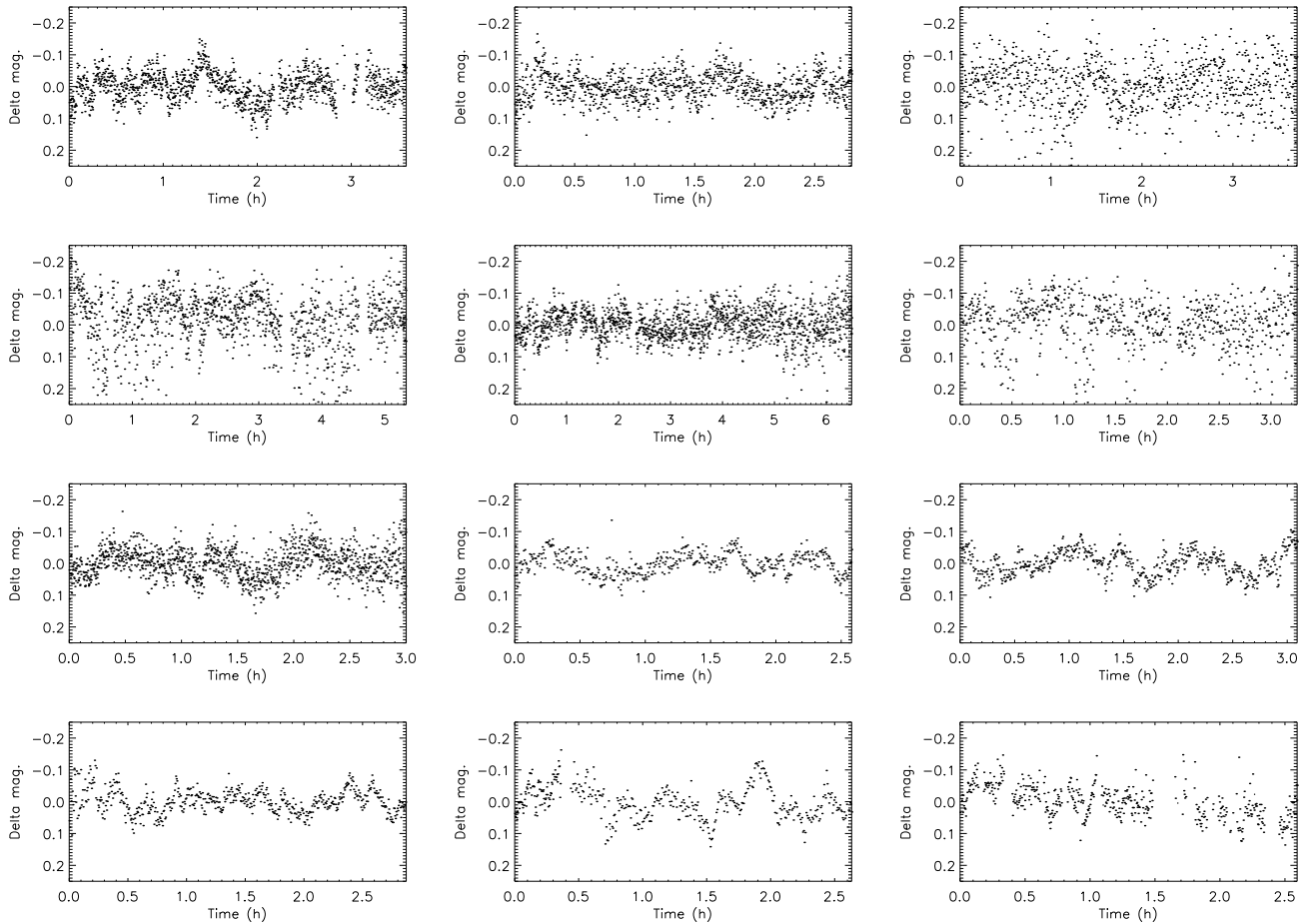}
\caption{The twelve sets of NOT white light light curves in chronological order (by rows). The strong,
almost periodic, flickering is present in most cases.}
\end{figure*}

\section[]{Observations}

We have obtained fast optical photometry of 4U0614+091 over several epochs spanning
over ten years in order to search for the orbital period, and attempt to unravel its remarkable temporal variability.
The observations have been carried out at the Nordic Optical Telescope, La Palma from 1998-2009 and consist mainly of
high time resolution white light CCD photometry time-series with a typical duration
of  3-4 hours per epoch (see detailed observing log in Table 1). Our data
are taken in white light using two different CCD detectors making their light curves not strictly comparable with each other due to the different QE properties. However, both the CCD's used are blue sensitive. Also, only differential photometry
was obtained each time, since we were mainly interested in looking for periodicities and also because 
there is no meaningful zero point calibration for white light ``instrumental magnitudes".  In order
to achieve fast readout the CCD images were sub-windowed and binned by a factor
of 2. The images have been bias-subtracted and flat-fielded in the normal manner. Light curves were extracted
using standard aperture photometry and a nearby star located 
22" WSW of the target was used as a comparison for the differential photometry. The resulting light curves show
strong and variable flickering and are plotted in Fig 1.   

In addition to the optical data, we use the RXTE ASM X-ray light curves and we have also revisited 
the original EXOSAT X-ray data published in Machin et al. (1990). EXOSAT ME light curve data products were 
extracted directly from the NASA-GSFC HEASARC database, and used as such.

\begin{table}
\caption{The observing log. All optical data were obtained in white light with a typical time resolution
of 10-15 s. The time resolution of archival EXOSAT ME data is 30 s.}
\begin{center}
\begin{tabular}{|c|c|c|c|}
\hline
\hline
Dataset \# & Site & Date  & Duration (h)  \\   
\hline
\hline
1 & NOT & 03-03-1998 &  3.6 \\
2 & NOT & 05-03-1998 &  2.8 \\
3 & NOT & 02-02-1999 &  3.7 \\
4 & NOT & 03-02-1999 &  5.3 \\
5 & NOT & 04-02-1999 &  6.5 \\
6 & NOT &05-02-1999 &  3.3 \\
7 & NOT & 01-11-1999 &  3.0 \\
8 & NOT & 05-01-2003 &  2.6 \\
9 & NOT & 06-01-2003 &  3.1 \\
10 & NOT & 07-01-2003 &  2.9 \\
11 & NOT & 21-11-2009 &  2.6 \\
12 & NOT & 23-11-2009 &  2.6 \\
\hline
13 & EXOSAT-ME & 12-02-1986 & 5.1 \\
\hline
\hline
\end{tabular}
\end{center}
\label{default}
\end{table}%

\begin{figure*}
\includegraphics[scale=1.0]{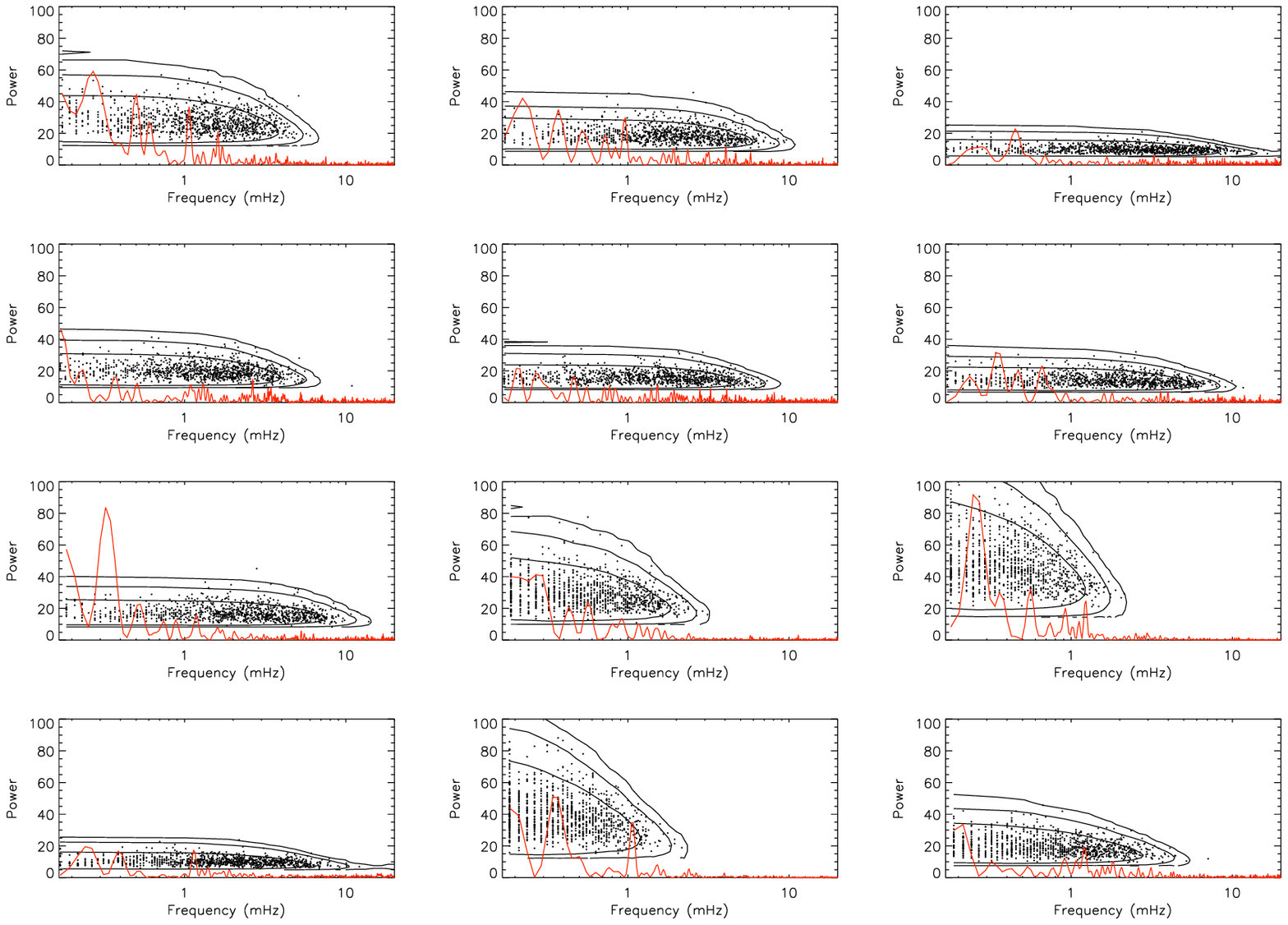}
\caption{The power spectra of different optical datasets, plotted together with the 95\%, 99\% and 99.9\&
red noise confidence contours from our Monte Carlo analysis. The individual points denote the 
distribution of highest peaks of a 1\% subsample of the 100000 Monte Carlo power spectra used to estimates the
confidence contours. The location of the highest peaks in the observed power spectra (red) are directly comparable to
the contours. }   
\end{figure*}

\section[]{The Time series analysis}

The time series analysis of our data has been performed in two stages. Firstly, we have
obtained Lomb-Scargle (Scargle 1982) power spectra of all the individual time series searching for any periodicities. The time resolution of our light curves varies,
but it is typically of the order of 10-20 seconds. Thus our analysis is sensitive to periods
down to about 40 sec. In the second phase of our analysis we have modelled the
variability in each of the light curves with a stochastic time series model and used those
as a basis for a Monte Carlo study of the noise properties of the light curves. This enabled 
us to obtain confidence limits on any peaks in the power spectra, even in the presence of
red (correlated) noise that is often present in accreting sources and evidenced by non-periodic
flickering behaviour. 

\subsection{Period analysis with presence of red noise}

The standard false alarm probability estimate from the Lomb-Scargle algorithm gives the 
significance of the highest peak in the power spectrum assuming that all the datapoints are 
independent. However, in the presence of correlated data (i.e. red noise) we will have to take
a different approach in order to properly estimate the significance of the peaks evident in the 
various power spectra. We have done this numerically by means of Monte Carlo simulations. 
Our approach consists of estimating the red noise component in the time domain using an autoregressive
time series model as a basis for our analysis. Autoregressive (AR) models (see Chatfield 1989 for time domain 
time series analysis) of different order can be fitted to equally spaced time series in order to estimate 
the red noise content. Once the correct AR model is chosen and fitted to the data, it can be used to 
simulate statistically equivalent data sets. By statistically equivalent we mean datasets that contain
the same variance and noise properties as the actual data. It is worth noting, that unlike the direct
bootstrapping approach, where the actual datapoints are picked in random order in order to simulate
synthetic light curves, the AR approach retains the correlations in between the neighbouring
data points, thus retaining the red noise properties of the dataset. 

In order to choose the right order for the AR modelling, we fitted the light curves with several AR
models with an increasing order (up to 6th order). The fitting itself is carried out by estimating 
the AR coefficients from the autocorrelations. We then examined the resulting correlations
(at different time lags) to see when the correlations become insignificant or level off (Chatfield 1989).
As a result we find that typically a second order autoregressive model i.e.

\begin{equation}
X_{t}=\alpha_{1}X_{t-1}+\alpha_{2}X_{t-2}+N(\sigma)
\end{equation}

where X$_{i}$ are the time series values and N($\sigma$) is the white noise component, is sufficient to 
describe the correlation properties in the data.  Thus a second order autoregressive
model (i.e. AR(2)) is used as a basis for the following significance estimates for the peaks seen in 
power spectra. The procedure for obtaining these estimates is as follows:

\begin{itemize}

\item Estimate the red noise model in the time domain by fitting an autoregressive
 time series model (we use a second order model i.e. AR(2)) to the data. 

\item Use the AR(2) model fit from above to generate 100,000 synthetic light curves with the same 
 time values and same noise properties as the real data.

\item Create Lomb-Scargle power spectra of the 100,000 synthetic light curves.

\item Identify the highest peak below 100 minutes in each of the synthetic power spectra and 
  measure its relative power (i.e. the power of the highest peak divided by the mean 
  power at all the periods below 100 min). 

\item Record the period/frequency of the highest peak together with its relative
 power in order to build a sample of 100,000 points in the period/frequency vs. power space.

\end{itemize}

\begin{figure}
\includegraphics[scale=0.5]{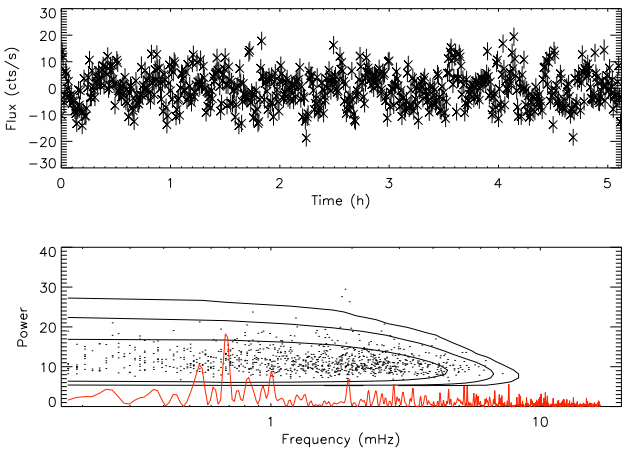}
\caption{The part of EXOSAT ME light curve exhibiting variability at a period 
of 24.5 minutes (top) and the power spectrum (red) with the 95\%, 99\% and 99.9\% confidence intervals
for the highest peak in power spectrum.}
\end{figure}

As a result, we will  get 100,000 frequency vs. maximum relative power pairs for each of our
real datasets. These are then plotted in the frequency vs. power space together with
the confidence contours that enclose 95\%, 99\% and 99.9\% of the cases (Fig 2). 
Finally, we express the periodogram power in terms of relative power in the same way
as was done above in step 4 for the synthetic data, and overplot it (in red, Fig 2). 
Now, if the highest peak in the 
observed periodogram falls outside the contours, the peak represents a likely real period
with significance indicated by the contours. Even if the contours form a closed loop, we are
naturally only interested if the highest peak in the power spectrum is {\it above} the contours
shown. The highest peak below the contours would mark a case, where there is only
white noise in the data, as the lowest horizontal sections of the contours represent the
case, where only white noise is present.

The results are quite striking when compared to the power spectra alone. Only in one dataset
out of 12 we find a peak well above the 99.9\% contour. That is the dataset showing 
a 50 min period (dataset \# 7, 3rd row, 1st column in Fig 2). The bottom row datasets in Fig 1 seem to
show clear periodic variability at periods around 15 min (1.1 mHz) in visual inspection.
Indeed, the period analysis shows peaks around that period. However, the red noise
analysis reveals that the peaks around 15 min (1.1 mHz) are at best 95\% significant 
(Fig 2, bottom row). This is also the case for all the other peaks in Fig 2 (apart from the
50 min period). This underlines the fact that, in the presence of significant amount of red noise,
neither the peak in the power spectrum or even the peak together with rather obvious visual
signal always guarantee a real periodicity! 

\subsection{Optical and X-ray anticorrelation}
Machin et al. (1990) obtained optical photometry simultaneously with the $\sim$ 14h EXOSAT
observation. This enabled them to conclude that the optical and X-ray emission are anticorrelated.
We have re-examined carefully the EXOSAT ME data (Fig 3) in search of flickering and/or short periods.
The EXOSAT ME observation in question consists of three parts. During the first part the X-ray 
flux is low and there is no evidence for any flickering. During the second part the X-ray flux is rising
and simultaneously the flickering is emerging. During the final part the X-ray flux is constantly
high and there appears to be a clear 'period' of flickering with a characteristic time scale of 24-25 min.
We have performed period analysis of the three parts of EXOSAT observations and whilst the 
first two do not show any evidence for any short period, the final part does seem to show the aforementioned 
behaviour. In order to evaluate the significance of the 24-25min period, we have performed red noise
analysis in the same manner as we did with the optical data above. The results are plotted in Fig 3 (bottom
panel). It is quite clear from our red noise analysis that the apparent periodicity can be attributed
to the presence of red noise (i.e. the significance of the 24-25min period is only $\sim$ 95\%). It is worth
noting though, that the EXOSAT period is very close to being half of the 51 min period reported by 
Shahbaz et al. (2008) and the 50 min period seen in one of our datasets.

\begin{figure}
\includegraphics[scale=0.5]{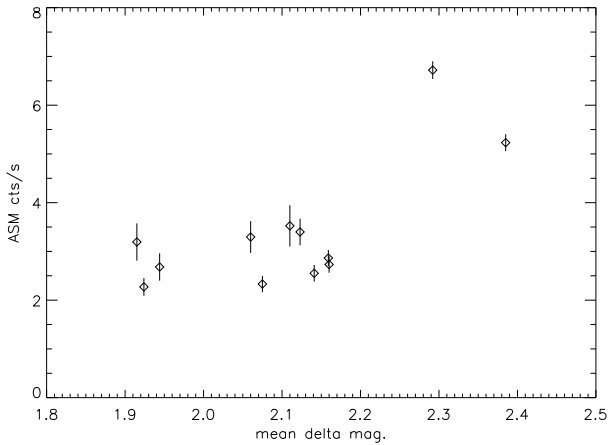}
\caption{The daily average RXTE-ASM count rate as a function of the mean relative white light magnitude from
our 12 optical datasets.}
\end{figure}

Given that Machin et al. (1990) reported X-ray vs. optical anticorrelation for the source, we have also studied 
the optical vs. X-ray anticorrelation by extracting the RXTE ASM daily average
count rates and comparing those with the average delta magnitudes in our optical data. We must interpret
the results with caution though, since even if all our optical data is in white light, the earlier observations 
are obtained with a different CCD detector than the latter ones. Both CCDs in question are blue sensitive and
AR coated though. The results from the optical vs- X-ray comparison are shown in Fig 4. The plot shows the 
optical magnitude vs. the X-ray flux.  

During most of the optical observations the X-ray flux is not correlated with the mean optical brightness. Only
during the two observations when the source is at it's faintest in the optical, can we see some evidence for
the optical vs. X-ray anticorrelation. The formal correlation coefficient for the whole set of twelve observations is 0.7
and the rank correlation test (Spearman's $\rho$) reveals that given the small number of points the apparent 
anticorrelation of fluxes is not very significant (10\% random chance probability) and if we remove the two rightmost
points from fig.4, there is no correlation left. However, given the
short term (within a day) variability of the RXTE ASM flux together with sampling effects and only 20\% changes 
in the X-ray flux level during the EXOSAT observations, a clear correlation would not be expected. The scatter
in ASM count rates (excluding the two rightmost points in Fig. 4) is 15\% (1$\sigma$) which could hide the
anticorrelation observed by Machin et al. (1990). Interestingly the two outlier points in Fig. 4 have peculiar values
both in their optical magnitude and their X-ray flux. These points correspond to the two first observations of our
dataset and thus could be caused by some longer term trend. 
 
Finally, we have examined the results of the AR(2) time series fits to the individual datasets in order to gain insight
into the underlying cause of variability. In particular, we have investigated whether the amount of red noise
(or correlated behaviour in the light curves) depends on the mean optical flux level. As noted earlier, 
the X-ray flickering in the EXOSAT data seemed to appear when the X-ray flux level reached its maximum.
However, in case of optical data we do not see any evidence for correlation in between the red noise level
(as estimated from the AR(2) fits) and the mean differential magnitude. Furthermore, the optical flux during
the only epoch of significant periodicity (dataset \#7) is not different from the mean optical flux level. The same
applies to the red noise properties.

\section{Conclusions}

We have analysed 12 sets of optical fast photometry of 4U0614+091 together with reanalysis of old EXOSAT ME
data. Despite the large amount of data, we do not have conclusive evidence for the orbital period of this 
peculiar system. However, in one case out of 13 we detect a strong periodic signal at around 50 min, which is
compatible with the tentative 48.5 min period reported by Nelemans (2006) and the 51.3 min period reported by 
Shahbaz et al. (2008) and cannot be explained by the presence of red noise. We can also conclude that none of our 
12 optical datasets show any evidence for eclipses (not even partial ones). This means that system inclinations above
75 degrees are very unlikely and there appears to be a controversy in between modelling the X-ray spectra (Schulz et al. 2010) 
and understanding the optical and X-ray light curves. The variety of optical light curves presented in this paper underline the multitude of variability present in this source. The fact that most of the seemingly significant (in terms of white noise alone) periodic signals can be explained using the red noise model does not mean that the variability is not real. 
It rather underlines the characteristic time scales at which the system exhibits nonperiodic variability such as flickering. 
We have tried to estimate whether these variability time scales would be correlated with different levels of X-ray or optical activity, 
but within our limited sample, this does not appear to be the case. However, one has to bear in mind that the RXTE ASM points we use are daily averages and the system is known to change it's optical and X-ray states at much shorter timescales. This, together 
with the small amplitude of detected X-ray level changes in the EXOSAT data (Machin et al., 1990)  certainly adds uncertainty 
to our optical vs. X-ray flux analysis. 

Finally, we would like to draw attention to the dangers that lie in the conventional period
analysis (i.e. chance estimation based on white noise only). The significance level analysis of any moderately significant 
peaks in the power spectra should always include the possible effects of red noise. This is particularly true for any accreting
sources where the flickering from the accretion flow will almost always introduce characteristic time scales in the data, that 
in turn show up as light curve data points that are not totally independent.

\section*{Acknowledgments}

The data presented here have been taken using ALFOSC, which is owned by the Instituto de Astrofisica de Andalucia (IAA) and operated at the Nordic Optical Telescope under agreement between IAA and the NBIfAFG of the Astronomical Observatory of Copenhagen. The authors gratefully acknowledge support from the
Academy of Finland. Finally, we would like to thank the anonymous referee for very constructive criticism that helped to improve the paper considerably.

\label{lastpage}

\end{document}